\documentclass[12pt]{article}
\usepackage{amsmath}
\usepackage{amssymb}
\usepackage{cite}
\newtheorem{Theorem}{Theorem}[section]

\title{\Large \textbf{Recursion operators and bi-Hamiltonian representations of cubic evolutionary (2+1)-dimensional systems}}
\author{\large\textbf{M. B. Sheftel$^1$ and D. Yaz{\i}c{\i}$^2$}\\
$^1$ Department of Physics, Bo\u{g}azi\c{c}i University, Bebek\\ 34342 Istanbul,  Turkey\\
$^2$ Department of Physics, Y{\i}ld{\i}z Technical University\\ Esenler, 34220 Istanbul, Turkey \\
E-mail: mikhail.sheftel@boun.edu.tr, yazici@yildiz.edu.tr}
\date{}

\begin{document}
\maketitle

\begin{abstract}
We construct all (2+1)-dimensional PDEs depending  only on 2nd-order derivatives of unknown which have the Euler-Lagrange form and determine the corresponding Lagrangians.
We convert these equations and their Lagrangians to two-component forms and find Hamiltonian representations of all these systems using Dirac's theory of constraints. We consider three-parameter integrable equations that are cubic in partial derivatives of the unknown applying our method of skew factorization of the symmetry condition. Lax pairs and recursion relations for symmetries are determined both for one-component and two-component forms.
For cubic three-parameter equations in the two-component form we obtain recursion operators in $2\times 2$ matrix form and bi-Hamiltonian representations, thus discovering three new bi-Hamiltonian (2+1) systems.
\end{abstract}

\section{Introduction}

We study integrability properties of (2+1)-dimensional second-order equations of the general form
\begin{equation}
  F = f - u_{tt}g = 0, \; (g\ne0)  \quad \iff \quad u_{tt} = \frac{f}{g}
 \label{F}
\end{equation}
where $f$ and $g$ are arbitrary smooth functions of $u_{t1},u_{t2},u_{11},u_{12},u_{22}$. Here $u=u(t,z_1,z_2)$, $u_{ti}=\partial^2u/\partial_t\partial_{z_i}$,
$u_{ij}=\partial^2u/\partial_{z_i}\partial_{z_j}$. Equations of the form \eqref{F} are equivalent to evolutionary two-component system
$u_t=v,\; v_t=G(v_1,v_2,u_{11},u_{12},u_{22})$.
Equations of this type arise in a wide range of applications, such as nonlinear physics, general relativity, differential geometry and integrable systems. Some examples are the Khokhlov-Zabolotskaya (dKP) equation in non-linear acoustics, the theory of Einstein-Weyl structures and the Boyer-Finley equation in self-dual gravity.

By integrability properties we mean Lax pairs, recursion operators and bi-Hamiltonian representations.
E. Ferapontov et al. studied integrability of equation \eqref{F} which was understood as the existence of infinitely many hydrodynamic reductions \cite{ferkhus3}. These authors showed the integrability condition to be equivalent to the property of equation \eqref{F} to be linearizable by contact transformations. However, this result does not provide a way to obtain explicitly such transformations and does not present a method to derive Lax pairs, recursion operators and bi-Hamiltonian representations of the considered equations.

Recently we studied the particular case $u_{tt}=f$, i.e. $g=1$ \cite{ShYa}, so here we concentrate on the case $g\not=1$. We are interested in all equations \eqref{F} which have the Euler-Lagrange form \cite{olv}. In other words, we consider the equations $u_{tt} = f/g$ which are Lagrangian or become Lagrangian after multiplication by $g$, an integrating factor of the variational calculus. According to Helmholtz conditions \cite{olv}, equation \eqref{F} is an Euler-Lagrange equation for a variational problem iff its Fr\'echet derivative operator is self-adjoint, $D_F^* = D_F$. We solve completely the latter equation and obtain explicitly all (2+1)-dimensional equations \eqref{F} which have the Euler-Lagrange form and their Lagrangians have also been constructed.

We utilize the method which we used earlier for constructing a degenerate Lagrangian for two-component evolutionary systems \cite{nns,sy,sym} and applying Dirac's theory of constraints \cite{dirac} in order to obtain Hamiltonian form of the system.

To discover integrable systems, we apply our modification \cite{ShYa} of A. Sergyeyev's method for constructing recursion operators \cite{Artur} which, as opposed to \cite{Artur}, does not require previous knowledge of Lax pairs. Our method consists in the skew factorization of symmetry condition for considered equation
and we explicitly demonstrate how to achieve this goal for a class of cubic three-parameter systems.

Composing a recursion operator in $2\times 2$ matrix form with the first Hamiltonian operator we obtain the second Hamiltonian operator and then find the corresponding Hamiltonian density, ending up with bi-Hamiltonian representations of the two-component systems.

We concentrate here on three-parameter systems that are cubic in the derivatives of unknown, the case which was not considered in our previous work \cite{ShYa}.

The paper is organized as follows. In Section \ref{sec-Lagrange}, we obtain all equations \eqref{F} that have the Euler-Lagrange form and derive their Lagrangians. In Subsection \ref{sec-2comp}, we convert our equations to two-component forms and construct degenerate Lagrangians for the two-component systems.
In Section \ref{sec-hamilton}, using the Dirac's theory of constraints we obtain the Hamiltonian operator $J_0$ and Hamiltonian density $H_1$ ending up with the Hamiltonian form of each of our systems. In Section \ref{sec-integr}, we give some comments on the integrability of multi-dimensional dispersionless equations and present an outline of the method of hydrodynamic reductions \cite{ferkhus3} by E. Ferapontov et. al., together with its illustration by the example of the dKP equation. Further we specify a subclass of cubic equations of the form \eqref{Fsol} for which we seek for recursion operators and bi-Hamiltonian structures.
In Section \ref{sec-symcond}, we show how converting the symmetry condition to a skew-factorized form we immediately extract Lax pair and recursion relations for symmetries. In three subsequent subsections, we obtain explicit results for cubic equations depending only on three parameters. In Section \ref{sec-biHam}, for reader's convenience we give a summary of notations used for bi-Hamiltonian system in the following sections and some remarks concerning the second Hamiltonian operator.

In Section \ref{RJ} we give a summary of new bi-Hamiltonian systems. In subsections \ref{sec_a5c4c5}, \ref{sec_c4c6c7} and \ref{sec_c5c7c8},
 we present recursion operators $R$ in a $2\times 2$ matrix form, the second Hamiltonian operators $J_1=RJ_0$ and Hamiltonian densities $H_0$ for two-component forms of the considered equations, thus obtaining bi-Hamiltonian representations of these systems. The results are given in the form of theorems without proofs.
The proofs are following in three subsequent sections \ref{sec-a5c4c5}, \ref{sec-c4c6c7} and \ref{sec-c5c7c8}.

\section{Lagrangian evolutionary equations in 2+1\\ dimensions}
 \setcounter{equation}{0}
\label{sec-Lagrange}

\begin{Theorem}
 The general Lagrangian equation of the type \eqref{F} has the following form
\begin{eqnarray}
&&\hspace*{-5.1pt} F = a_1\!\left\{u_{tt}(u_{11}u_{22} - u_{12}^2) - u_{t1}(u_{t1}u_{22} - u_{t2}u_{12}) + u_{t2}(u_{t1}u_{12} - u_{t2}u_{11})\right\}\nonumber\\
&&\hspace*{-5.1pt} + a_2(u_{tt}u_{11} - u_{t1}^2) + a_3(u_{tt}u_{12} - u_{t1}u_{t2}) + a_4(u_{tt}u_{22} - u_{t2}^2) + a_5u_{tt} \nonumber\\
&&\hspace*{-5.1pt} + c_1(u_{t1}u_{12} - u_{t2}u_{11}) + c_2(u_{t1}u_{22} - u_{t2}u_{12}) + c_3(u_{11}u_{22} - u_{12}^2)\nonumber\\
&&\hspace*{-5.1pt} + c_4u_{t1} + c_5u_{t2} + c_6u_{11} + c_7u_{12} + c_8u_{22} + c_9 = 0.
 \label{Fsol}
\end{eqnarray}
\end{Theorem}

{\bf Proof:}
The Fr\'echet derivative operator is defined as $D_F = \sum_J F_J[u]D_J$, where $J=(u,u_t,u_1,u_2,u_{tt},u_{t1},u_{t2},u_{11},u_{12},u_{22})$ is a multi-index and $D_J$ is the corresponding total derivative \cite{olv}, or explicitly
\begin{eqnarray*}
&& D_F = - gD_t^2 + (f_{u_{t1}} - u_{tt}g_{u_{t1}})D_tD_1 + (f_{u_{t2}} - u_{tt}g_{u_{t2}})D_tD_2 \\
&& + (f_{u_{11}} - u_{tt}g_{u_{11}})D_1^2 + (f_{u_{12}} - u_{tt}g_{u_{12}})D_1D_2 + (f_{u_{22}} - u_{tt}g_{u_{22}})D_2^2,
\end{eqnarray*}
\begin{eqnarray*}
&& D_F^* = - D_t^2g + D_tD_1(f_{u_{t1}} - u_{tt}g_{u_{t1}}) + D_tD_2(f_{u_{t2}} - u_{tt}g_{u_{t2}}) \\
&& + D_1^2(f_{u_{11}} - u_{tt}g_{u_{11}}) + D_1D_2(f_{u_{12}} - u_{tt}g_{u_{12}}) + D_2^2(f_{u_{22}} - u_{tt}g_{u_{22}}).
\end{eqnarray*}
The Helmholtz condition implies
\[-2D_t[g] + D_1[f_{u_{t1}} - u_{tt}g_{u_{t1}}] + D_2[f_{u_{t2}} - u_{tt}g_{u_{t2}}] = 0\]
where the square brackets denote taking the values of differential operators. Splitting this equation in third derivatives we obtain
$g_{u_{t1}} = g_{u_{t2}} = 0$ and the last equation simplifies to the following one
\begin{equation}
  -2D_t[g] + D_1[f_{u_{t1}}] + D_2[f_{u_{t2}}] = 0.
 \label{1}
\end{equation}
In a similar way we obtain from the Helmholtz condition three more equations
\begin{eqnarray}
&& D_t[f_{u_{t1}}] + 2 D_1[f_{u_{11}}] - 2u_{tt}D_1[g_{u_{11}}] - 2u_{tt1}g_{u_{11}}\nonumber\\
&& + D_2[f_{u_{12}}] - u_{tt}D_2[g_{u_{12}}] - u_{tt2}g_{u_{12}} = 0
 \label{2}
\end{eqnarray}
\begin{eqnarray}
&& D_t[f_{u_{t2}}] + D_1[f_{u_{12}}] - u_{tt}D_1[g_{u_{12}}] - u_{tt1}g_{u_{12}}\nonumber\\
&& + 2D_2[f_{u_{22}}] - 2u_{tt}D_2[g_{u_{22}}] - 2u_{tt2}g_{u_{22}} = 0
 \label{3}
\end{eqnarray}
\begin{eqnarray}
&& - D_t^2[g] + D_tD_1[f_{u_{t1}}] + D_tD_2[f_{u_{t2}}]\nonumber\\
&& + D_1^2[f_{u_{11}} - u_{tt}g_{u_{11}}] + D_2^2[f_{u_{22}} - u_{tt}g_{u_{22}}] = 0.
 \label{4}
\end{eqnarray}
The general solution of these equations results in the formula \eqref{Fsol}.
\hfill $\blacksquare$

\begin{Theorem}
Lagrangian for equation \eqref{Fsol} reads
\begin{eqnarray}
&&\hspace*{-5.4pt} L = a_1\!\frac{u}{4}\!\{u_{tt}(u_{11}u_{22} - u_{12}^2) - u_{t1}(u_{t1}u_{22} - u_{t2}u_{12}) + u_{t2}(u_{t1}u_{12} - u_{t2}u_{11})\}\nonumber\\
&&\hspace*{-5.4pt} + \frac{u}{3}\{a_2(u_{tt}u_{11} - u_{t1}^2) + a_3(u_{tt}u_{12} - u_{t1}u_{t2}) + a_4(u_{tt}u_{22} - u_{t2}^2)\nonumber\\
&&\hspace*{-5.4pt} + c_1(u_{t1}u_{12} - u_{t2}u_{11}) + c_2(u_{t1}u_{22} - u_{t2}u_{12}) + c_3(u_{11}u_{22} - u_{12}^2)\}\nonumber\\
&&\hspace*{-5.4pt} + \frac{u}{2}(a_5u_{tt} + c_4u_{t1} + c_5u_{t2} + c_6u_{11} + c_7u_{12} + c_8u_{22}) + c_9u .
 \label{L}
\end{eqnarray}
\end{Theorem}

{\bf Proof:}
The homotopy formula \cite{olv} $L[u] = \int_0^1 uF(\lambda u)\,d\lambda$, applied to $F$ in \eqref{Fsol}, yields the result \eqref{L}.
\hfill $\blacksquare$

\subsection{Two-component form}
\label{sec-2comp}

The two-component form of our equation $F=0$ with $F$ given in \eqref{Fsol} reads
\begin{eqnarray}
&& u_t = v,\nonumber\\
&& v_t = \frac{1}{\Box}\{a_1(v_1^2u_{22} - 2v_1v_2u_{12} + v_2^2u_{11}) + a_2v_1^2 + a_3v_1v_2 + a_4v_2^2\nonumber\\
&& + c_1(v_2u_{11} - v_1u_{12}) + c_2(v_2u_{12} - v_1u_{22}) + c_3(u_{12}^2 - u_{11}u_{22})\nonumber\\
&& - c_4v_1 - c_5v_2 - c_6u_{11} - c_7u_{12} - c_8u_{22} - c_9\}
 \label{2comp}
\end{eqnarray}
where
\begin{equation}
\Box = a_1(u_{11}u_{22} - u_{12}^2) + a_2u_{11} + a_3u_{12} + a_4u_{22} + a_5.
\label{Delt}
\end{equation}
\begin{Theorem}
Lagrangian for the two-component form \eqref{2comp} of our system has the form
\begin{eqnarray}
&& L = \left(u_tv - \frac{v^2}{2}\right)\!\{a_1(u_{11}u_{22} - u_{12}^2) + a_2u_{11} + a_3u_{12} + a_4u_{22} + a_5\}\nonumber\\
&& + u_t(c_1u_1 - c_2u_2)u_{12} + \frac{u_t}{2}(c_4u_1 + c_5u_2) - \frac{c_3}{3}u(u_{11}u_{22} - u_{12}^2)\nonumber\\
&& - \frac{u}{2}(c_6u_{11} + c_7u_{12} + c_8u_{22}) - c_9u.
 \label{L2}
\end{eqnarray}
\end{Theorem}

{\bf Proof:}
We modify the Lagrangian $L$ given in \eqref{L} and skip some total derivative terms
to obtain the result \eqref{L2}.
\hfill $\blacksquare$

\section{Hamiltonian representation}
\setcounter{equation}{0}
\label{sec-hamilton}

\begin{Theorem}
The system \eqref{2comp} can be presented in the Hamiltonian form
\begin{equation}
\left(
\begin{array}{c}
 u_t\\
 v_t
\end{array}
\right) = J_0
\left(
\begin{array}{c}
 \delta_u H_1\\
 \delta_v H_1
\end{array}
\right)
  \label{hamiltform1}
  \end{equation}
where $\delta_u$ and $\delta_v$ are Euler-Lagrange operators \cite{olv} with respect to $u$ and $v$, respectively.
Here $J_0$ is the Hamiltonian operator, which determines the structure of the Poisson bracket, defined as
  \begin{equation}
    J_0 =   \left(
    \begin{array}{rc}
   0 & - K_{21}^{-1}\\
  K_{12}^{-1} & K_{12}^{-1}K_{11}K_{12}^{-1}
      \end{array}
    \right) =
    \left(
    \begin{array}{rc}
   0 & \frac{1}{\Box}\\
  -\frac{1}{\Box} & \frac{1}{\Box}K_{11}\frac{1}{\Box}
    \end{array}
    \right).
\label{hamilton1}
\end{equation}
where
\begin{eqnarray}
&& K_{11} = a_1\{2(v_1u_{22} - v_2u_{12})D_1 + 2(v_2u_{11} - v_1u_{12})D_2\nonumber\\
&&  + v_{11}u_{22} + v_{22}u_{11} - 2v_{12}u_{12}\}\nonumber\\
&& + a_2(2v_1D_1 + v_{11}) + a_3(v_2D_1 + v_1D_2 + v_{12}) + a_4(2v_2D_2 + v_{22})\nonumber\\
&& + c_1(u_{11}D_2 - u_{12}D_1) + c_2(u_{12}D_2 - u_{22}D_1) - c_4D_1 - c_5D_2
 \label{K11'}
\end{eqnarray}
whereas $H_1$ is the corresponding Hamiltonian density defined as
\begin{eqnarray}
&& H_1 = \frac{1}{2}v^2\Box + \frac{c_3}{3}u(u_{11}u_{22} - u_{12}^2) + \frac{u}{2}(c_6u_{11} + c_7u_{12} + c_8u_{22})\nonumber\\
&& + c_9u.
 \label{H1}
\end{eqnarray}
\end{Theorem}

{\bf Proof:}
We define canonical momenta
\begin{eqnarray}
&& \pi_u = \frac{\partial L}{\partial u_t} = v\{a_1(u_{11}u_{22} - u_{12}^2) + a_2u_{11} + a_3u_{12} + a_4u_{22} + a_5\}\nonumber\\
&& + (c_1u_1 - c_2u_2)u_{12} + \frac{1}{2}(c_4u_1 + c_5u_2),\quad \pi_v = \frac{\partial L}{\partial v_t} = 0
 \label{piu}
\end{eqnarray}
which satisfy canonical Poisson brackets
\[[\pi_i(z),u^k(z')] = \delta_i^k\delta(z-z')\]
where $u^1=u$, $u^2=v$, $z=(z_1,z_2)$. The only nonzero Poisson bracket is $[\pi_u,u] = \delta(z_1-z_1')\delta(z_2-z_2')$.

The Lagrangian \eqref{L2} is degenerate because the momenta cannot be inverted for the velocities. Therefore, following the Dirac's theory of constraints \cite{dirac}, we impose \eqref{piu} as constraints
\begin{eqnarray*}
&& \Phi_u = \pi_u  - v\{a_1(u_{11}u_{22} - u_{12}^2) + a_2u_{11} + a_3u_{12} + a_4u_{22} + a_5\}\\
&& - (c_1u_1 - c_2u_2)u_{12} - \frac{1}{2}(c_4u_1 + c_5u_2),\\
&& \Phi_v = \pi_v
\end{eqnarray*}
and calculate Poisson brackets for the constraints
\begin{eqnarray*}
 && K_{11} = [\Phi_u(z_1,z_2),\Phi_{u'}(z'_1,z'_2)], \quad K_{12} = [\Phi_u(z_1,z_2),\Phi_{v'}(z'_1,z'_2)] \nonumber\\
 && K_{21} = [\Phi_v(z_1,z_2),\Phi_{u'}(z'_1,z'_2)],\quad K_{22} = [\Phi_v(z_1,z_2),\Phi_{v'}(z'_1,z'_2)].
% \label{constr}
\end{eqnarray*}
We obtain the following matrix of Poisson brackets
\begin{equation}
  K = \left(
  \begin{array}{cc}
     K_{11} & - \Box \\
       \Box      &  0
  \end{array}
  \right)
\label{K}
\end{equation}
where $K_{11}$ is defined in \eqref{K11'}
or in a skew-symmetric form
\begin{eqnarray}
&& K_{11} = a_1\{D_1(v_1u_{22} - v_2u_{12}) + (v_1u_{22} - v_2u_{12})D_1\nonumber\\
&& + D_2(v_2u_{11} - v_1u_{12}) + (v_2u_{11} - v_1u_{12})D_2\}\nonumber\\
&& + a_2(D_1v_1 + v_1D_1) + a_3(D_1v_2 + v_2D_1 + D_2v_1 + v_1D_2)\nonumber\\
&& + a_4(D_2v_2 + v_2D_2) + c_1 (D_2u_{11} - D_1u_{12}) + c_2(D_2u_{12} - D_1u_{22})\nonumber\\
&& - c_4D_1 - c_5D_2.
 \label{K11}
\end{eqnarray}

The Hamiltonian operator, which determines the structure of the Poisson bracket, is the inverse to the symplectic operator $J_0 = K^{-1}$

Here
\begin{equation}
J_0^{22} = \frac{1}{\Box}K_{11}\frac{1}{\Box}
 \label{J022}
\end{equation}
with $K_{11}$ defined by \eqref{K11}. More precisely,
operator $J_0$ is Hamiltonian if and only if its inverse $K$ is symplectic \cite{ff}, which means that the volume integral $\Omega = \iiint_V\omega dV$ of $\omega = (1/2)du^i\wedge K_{ij}du^j$ should be a symplectic form, i.e. at appropriate boundary conditions $d\omega = 0$ modulo total divergence. Another way of formulation is to say that the vertical differential of $\omega$ should vanish \cite{Kras}.
In $\omega$ summations over $i,j$ run from 1 to 2 and $u^1 = u,\; u^2 = v$. Using \eqref{K}, we obtain
\begin{equation}
\omega = \frac{1}{2}du\wedge K_{11}du - \Box du\wedge dv.
\label{omega}
\end{equation}
Using here $K_{11}$ from \eqref{K11}, taking exterior derivative of \eqref{omega} and skipping total divergence terms, we have checked that $d\omega=0$ which proves that operator $K$ is symplectic and hence $J_0$ defined in \eqref{hamilton1} is indeed a Hamiltonian operator.

The corresponding Hamiltonian density $H_1$ is defined by $H_1 = u_t\pi_u + v_t\pi_v - L$ with the result \eqref{H1}.

The Hamiltonian form of this system is determined by \eqref{hamiltform1}.
\hfill $\blacksquare$

\section{Some comments on integrability}
\setcounter{equation}{0}
\label{sec-integr}

There are various approaches to integrability of multi-dimensional systems. The most traditional one uses the Lax pair and the inverse scattering transform.
However, there are integrable equations to which this method cannot be applied. The well-known example is the dispersionless Kadomtsev-Petviashvili (dKP) equation (in our notation $y\mapsto t, x\mapsto z_1, t\mapsto z_2$)
\begin{equation}
u_{yy} = u_{xt} - \frac{1}{2}u_{xx}^2.
\label{dkp}
\end{equation}
E. Ferapontov et. al. \cite{ferkhus3,ferkhus1} use more general definition of integrability which requires the existence of infinitely many hydrodynamic reductions. According to this method, one decouples a three-dimensional PDE, like dKP, into a pair of commuting $n$-component (1+1)-dimensional systems of hydrodynamic type
\begin{equation}
 R^i_t = \lambda^i(R)R^i_x,\quad R^i_y = \mu^i(R)R^i_x
\label{hydrotype}
\end{equation}
where the characteristic speeds $\lambda^i$ and $\mu^i$ satisfy the compatibility conditions
\[\frac{\partial_j\lambda^i}{\lambda^j-\lambda^i} = \frac{\partial_j\mu^i}{\mu^j-\mu^i} ,\qquad i\ne j.\]
Here $\partial_j = \partial_{R^j}$. For the example of dKP, the authors in \cite{ferkhus3} use its quasilinear representation
\begin{equation}
a_y=b_x,\quad a_t=p_x,\quad b_t=p_y,\quad b_y = \left(p - \frac{1}{2}a^2\right)_x
\label{quasilin}
\end{equation}
with the notation $a=u_{xx},\; b=u_{xy},\; p=u_{xt}$. Seek multi-phase solutions $a=a(R), b=b(R), p=p(R)$ with $R=(R^1,\ldots,R^n)$
where the phases $R^i(x,y,t)$ satisfy equations \eqref{hydrotype}. Then \eqref{quasilin} implies
\[\partial_ib = \mu^i\partial_ia,\quad \partial_ip = \lambda^i\partial_ia,\quad \lambda^i = a + (\mu^i)^2\]
with the compatibility conditions resulting in the Gibbons-Tsarev system for $a(R)$ and $\mu^i(R)$
\begin{equation}
 \partial_j\mu^i =  \frac{\partial_ja}{\mu^j-\mu^i},\quad \partial_i\partial_ja =  \frac{2\partial_ia\partial_ja}{(\mu^j-\mu^i)^2},\qquad i\ne j,\quad
 i,j = 1,\ldots, n.
 \label{GibTsar}
\end{equation}
This system is in involution and its general solution depends on $n$ arbitrary functions of one variable. Thus, the dKP equation possesses infinitely many
$n$-component reductions parameterized by $n$ arbitrary functions of one variable. This result suggests the definition \cite{ferkhus3}.

\textbf{Definition}
An equation of the form \eqref{F} is said to be \textit{integrable} if, for any $n$,  \textit{it possesses infinitely many $n$-component hydrodynamic reductions
parameterized by $n$ arbitrary functions of one variable.}

Our concept of integrability implicates the existence of Lax pairs, recursion operators and bi-Hamiltonian representations. However, we note that the form of equation \eqref{Fsol} is exactly the same as obtained by E. Ferapontov et. al. (equation (21) in \cite{ferkhus3})
with an appropriate change of notation.

Thus,  the requirement that the equation \eqref{F} should have the Euler-Lagrangian form yields the same form of equation as the one obtained by the method of hydrodynamic reductions.

E. Ferapontov et. al. have derived in \cite{ferkhus3} an integrability condition for the symplectic Monge--Amp\`ere equation \eqref{Fsol}, the integrability meaning that the equation \eqref{Fsol} admits infinitely many hydrodynamic reductions \cite{ferkhus1}. The integrability turns out to be equivalent to linearizability of equation \eqref{Fsol} though it does not show the way how the linearizing transformation could be explicitly found. It was shown in \cite{ferkhus3} that one can distinguish two cases for equation \eqref{Fsol} together with its integrability condition. \\
\textit{Case I}: $a_1 = 1$,\; $a_2=a_3=a_4=0$,\; $c_1=c_2=c_3=0$,\\
Equation simplifies to the following one
\begin{eqnarray}
&& det\left[\begin{array}{ccc}
 u_{tt} & u_{t1} & u_{t2} \\
 u_{1t} & u_{11} & u_{12} \\
 u_{2t} & u_{21} & u_{22}
\end{array}\right] + a_5u_{tt} + c_4u_{t1} + c_5u_{t2} \nonumber\\
&& + c_6u_{11} + c_7u_{12} + c_8u_{22} + c_9 = 0.
 \label{I}
\end{eqnarray}
Integrability condition becomes
\begin{equation}
4a_5c_6c_8 + c_4c_5c_7 + c_9^2 - (c_5^2c_6 + c_4^2c_8 + c_7^2a_5) = 0.
 \label{intI}
\end{equation}
\textit{Case II}: $\varepsilon\equiv a_1 = 0$.\: There are no further simplifications.\\
In our earlier paper \cite{ShYa} we have studied the subcase of the case II with
$a_2=a_3=a_4=0$, $a_5 = -1$.

Here we concentrate on the case $I$ where the equation has the form \eqref{I}.

\section{Symmetry condition and integrability}
\setcounter{equation}{0}
\label{sec-symcond}

Our method is based on the presentation of the symmetry condition of the considered equation in the skew-factorized form.

Symmetry condition is the differential compatibility condition of equation \eqref{Fsol} and the Lie equation $u_\tau = \varphi$, where
$\varphi$ is the symmetry characteristic and $\tau$ is the group parameter. It has the form of Fr\'echet derivative (linearization) of equation \eqref{Fsol},
 $D_\tau F = M[\varphi] = 0$,  with the operator $M$ of the symmetry condition given by
\begin{eqnarray}
&& M = a_1 \left[(u_{11}u_{22} - u_{12}^2)D_t^2 + u_{tt}(u_{22}D_1^2 + u_{11}D_2^2 - 2u_{12}D_1D_2)\right.\nonumber\\
&&\left. + 2u_{t1}(u_{12}D_tD_2 - u_{22}D_tD_1) + 2u_{t2}(u_{12}D_tD_1 - u_{11}D_tD_2)\right.\nonumber\\
&&\left. + 2u_{t1}u_{t2}D_1D_2 - u_{t2}^2D_1^2 - u_{t1}^2D_2^2\right] + a_2(u_{11}D_t^2 + u_{tt}D_1^2 - 2u_{t1}D_tD_1)\nonumber\\
&& + a_3(u_{12}D_t^2 + u_{tt}D_1D_2 - u_{t2}D_tD_1 - u_{t1}D_tD_2)\nonumber\\
&& + a_4(u_{22}D_t^2 + u_{tt}D_2^2 - 2u_{t2}D_tD_2) + a_5D_t^2\nonumber\\
&& + c_1(u_{12}D_tD_1 + u_{t1}D_1D_2 - u_{11}D_tD_2 - u_{t2}D_1^2)\nonumber\\
&& + c_2(u_{22}D_tD_1 + u_{t1}D_2^2 - u_{12}D_tD_2 - u_{t2}D_1D_2)\nonumber\\
&& + c_3(u_{22}D_1^2 + u_{11}D_2^2 - 2u_{12}D_1D_2) + c_4D_tD_1 + c_5D_tD_2 + c_6D_1^2\nonumber\\
&& + c_7D_1D_2 + c_8D_2^2.
 \label{sym}
\end{eqnarray}

In case $I$, the symmetry condition becomes $M[\varphi] = 0$ with $M$ defined as
 \begin{eqnarray}
&& M = \left[(u_{11}u_{22} - u_{12}^2)D_t^2 + u_{tt}(u_{22}D_1^2 + u_{11}D_2^2 - 2u_{12}D_1D_2)\right.\nonumber\\
&&\left. + 2u_{t1}(u_{12}D_tD_2 - u_{22}D_tD_1) + 2u_{t2}(u_{12}D_tD_1 - u_{11}D_tD_2)\right.\nonumber\\
&&\left. + 2u_{t1}u_{t2}D_1D_2 - u_{t2}^2D_1^2 - u_{t1}^2D_2^2\right] + a_5D_t^2\nonumber\\
&& + c_4D_tD_1 + c_5D_tD_2 + c_6D_1^2 + c_7D_1D_2 + c_8D_2^2.
 \label{symI}
\end{eqnarray}

We apply our modification \cite{ShYa} of A. Sergyeyev's method for constructing recursion operators \cite{Artur} which does not require previous knowledge of the Lax pair. For integrable equation \eqref{I}, the symmetry condition $M[\varphi] = 0$,  with the operator $M$ given by \eqref{symI}, should be presented in the skew-factorized form
\begin{equation}
 (A_1B_2 - A_2B_1)\varphi = 0
 \label{factor}
\end{equation}
where the commutator relations
\begin{equation}
  [A_1,A_2] = 0,\quad [A_1,B_2] - [A_2,B_1] = 0,\quad [B_1,B_2] = 0
 \label{commut}
\end{equation}
should be satisfied on solutions of equation \eqref{I}. It immediately follows that the following two operators also commute on solutions
\begin{equation}
   X_1 = \lambda A_1 + B_1,\quad X_2 = \lambda A_2 + B_2, \qquad [X_1,X_2] = 0
  \label{Lax}
\end{equation}
and therefore constitute Lax representation for equation \eqref{I}  with $\lambda$ being a spectral parameter.

Symmetry condition in the form \eqref{factor}
also provides the recursion relations for symmetries
\begin{equation}
  A_1\tilde{\varphi} = B_1\varphi,\quad A_2\tilde{\varphi} = B_2\varphi
 \label{recurs}
\end{equation}
where $\tilde{\varphi}$ satisfies symmetry condition $M[\tilde{\varphi}] = 0$ if and only if $\varphi$ is also a symmetry, $M[\varphi] = 0$. The latter claim follows from the consequences of relations \eqref{commut}
\begin{eqnarray*}
&& (A_1B_2 - A_2B_1)\varphi = [A_1,A_2]\tilde{\varphi} = 0,\; (A_1B_2 - A_2B_1)\tilde{\varphi} = [B_2,B_1]\varphi = 0.
\end{eqnarray*}

It is convenient to introduce first-order differential operators
\begin{eqnarray}
&& M_{12} = (u_{11}u_{22} - u_{12}^2)D_t - (u_{t1}u_{22} - u_{t2}u_{12})D_1 + (u_{t1}u_{12} - u_{t2}u_{11})D_2\nonumber\\
&& M_{2t} = - (u_{t1}u_{22} - u_{t2}u_{12})D_t + (u_{tt}u_{22} - u_{t2}^2)D_1 - (u_{tt}u_{12} - u_{t1}u_{t2})D_2\nonumber\\
&& M_{t1} = (u_{t1}u_{12} - u_{t2}u_{11})D_t - (u_{tt}u_{12} - u_{t1}u_{t2})D_1 + (u_{tt}u_{11} - u_{t1}^2)D_2.\nonumber\\
 \label{Mij}
\end{eqnarray}
Operator $M$ of symmetry condition \eqref{symI} becomes
\begin{eqnarray}
&& M = M_{12}D_t + M_{2t}D_1 + M_{t1}D_2 + a_5D_t^2\nonumber\\
&& + c_4D_tD_1 + c_5D_tD_2 + c_6D_1^2 + c_7D_1D_2 + c_8D_2^2.
 \label{M}
\end{eqnarray}
We note the identities
\begin{eqnarray}
&& u_{tt}M_{12} + u_{t1}M_{2t} + u_{t2}M_{t1} = \Delta D_t,\nonumber\\
&& u_{t1}M_{12} + u_{11}M_{2t} + u_{12}M_{t1} = \Delta D_1,
 \label{ident}
\\ && u_{t2}M_{12} + u_{12}M_{2t} + u_{22}M_{t1} = \Delta D_2,\nonumber
\end{eqnarray}
\begin{equation}
\Delta = det\left[\begin{array}{ccc}
 u_{tt} & u_{t1} & u_{t2} \\
 u_{1t} & u_{11} & u_{12} \\
 u_{2t} & u_{21} & u_{22}
\end{array}\right]
 \label{Del}
\end{equation}
which can be replaced due to equation \eqref{I} by the expression
\begin{equation}
\Delta = - (a_5u_{tt} + c_4u_{t1} + c_5u_{t2} + c_6u_{11} + c_7u_{12} + c_8u_{22} + c_9).
 \label{eqI}
\end{equation}

In the following, we will use obvious symmetries of the equation \eqref{I} generated by $X=\partial/\partial_t$, $X=\partial/\partial_1$, $X=\partial/\partial_2$ with the symmetry characteristics $\varphi=u_t$, $\varphi=u_1$, $\varphi=u_2$, respectively, which identically satisfy symmetry condition $M[\varphi] = 0$ with $M$ defined in \eqref{M}.

\subsection{$(a_5c_4c_5)$-parameter system}

We consider here the three-parameter equation obtained by setting $c_6 = c_7 = c_8 = c_9 = 0$ in \eqref{eqI}
\begin{equation}
  \Delta + a_5u_{tt} + c_4u_{t1} + c_5u_{t2} = 0
 \label{a5c4c5}
\end{equation}
where $\Delta$ is defined in \eqref{Del}.

\begin{Theorem}
 Equation \eqref{a5c4c5} is integrable with the Lax pair
\begin{eqnarray}
& X_1 = \displaystyle\frac{1}{u_{tt}}(\lambda (M_{2t} +c_4D_t) + u_{t2}D_t - u_{tt}D_2) \nonumber\\
& X_2 = \displaystyle\frac{1}{u_{tt}}(\lambda (M_{t1} + c_5D_t) + u_{tt}D_1 - u_{t1}D_t).
 \label{laxa5}
\end{eqnarray}
 and the recursions for symmetries
\begin{equation}
(M_{2t} + c_4D_t)\tilde{\varphi} = (u_{t2}D_t - u_{tt}D_2)\varphi,\quad (M_{t1} + c_5D_t)\tilde{\varphi} = (u_{tt}D_1 - u_{t1}D_t)\varphi.
 \label{rec_a5}
\end{equation}
\end{Theorem}

{\bf Proof:}
The symmetry condition for equation \eqref{a5c4c5} becomes
\begin{equation}
 M[\varphi] = \{M_{12}D_t + M_{2t}D_1 + M_{t1}D_2 + a_5D_t^2 + c_4D_tD_1 + c_5D_tD_2\}\varphi  = 0.
 \label{syma5}
\end{equation}
It is identically satisfied by $\varphi = u_t$
\begin{equation}
  M_{12}[u_{tt}] + M_{2t}[u_{t1}] + M_{t1}[u_{t2}] + a_5u_{ttt} + c_4u_{tt1} + c_5u_{tt2} = 0
 \label{a5ut}
\end{equation}
where the square brackets denote values of operators. We combine \eqref{a5ut} with the first equation from \eqref{ident}
to obtain
\begin{equation}
  M_{12}u_{tt} + M_{2t}u_{t1} + M_{t1}u_{t2} + D_t(a_5u_{tt} + c_4u_{t1} + c_5u_{t2}) = 0.
 \label{Mutt}
\end{equation}
We use \eqref{Mutt} in the identity transformation
\begin{eqnarray}
&& M_{12}D_t = M_{12}u_{tt}\frac{1}{u_{tt}}D_t\nonumber\\
&& = - \{M_{2t}u_{t1} + M_{t1}u_{t2} + D_t(a_5u_{tt} + c_4u_{t1} + c_5u_{t2})\}\frac{1}{u_{tt}}D_t.
\label{M12Dt}
\end{eqnarray}
Applying this to the symmetry condition \eqref{syma5} we transform it to the skew-factorized form \eqref{factor}
\begin{equation}
  \left\{(M_{2t} + c_4D_t)\left(D_1 - \frac{u_{t1}}{u_{tt}}D_t\right) + (M_{t1} + c_5D_t)\left(D_2 - \frac{u_{t2}}{u_{tt}}D_t\right)\right\}\varphi = 0.
 \label{skewa5}
\end{equation}
We adopt the definitions
\begin{eqnarray}
& A_1 = \displaystyle\frac{1}{u_{tt}}(M_{2t} + c_4D_t),\quad A_2 = \frac{1}{u_{tt}}(M_{t1} + c_5D_t)\nonumber\\
& B_1 = - D_2 \displaystyle + \frac{u_{t2}}{u_{tt}}D_t,\quad B_2 = D_1 - \frac{u_{t1}}{u_{tt}}D_t
 \label{a5AiBi}
\end{eqnarray}
so that \eqref{skewa5} takes the skew-factorized form \eqref{factor}.
A straightforward check shows that all the integrability conditions \eqref{commut} are identically satisfied on solutions of equation \eqref{a5c4c5}, together with their immediate consequences for the Lax pair \eqref{Lax} and  recursion relations for symmetries \eqref{recurs}.

If we choose $B_i$ to be the second factors in \eqref{skewa5}, then $A_i$ are defined by \eqref{skewa5} only up to a common factor.
It is interesting to note that choosing this factor to be $1/u_{tt}$, the same as involved in $B_i$,
we obtain as a consequence that all the operators $A_i$ and $B_i$ automatically satisfy all the conditions \eqref{commut}. A similar property holds for three-parameter equations given below.

The Lax pair is constituted by the operators \eqref{laxa5}.

The recursions for symmetries are given by \eqref{rec_a5}.
\hfill $\blacksquare$

\subsection{$(c_4c_6c_7)$-parameter system}

Here we consider the three-parameter equation obtained by setting $a_5 = c_5 = c_8 = c_9 = 0$ in \eqref{eqI}
\begin{equation}
  \Delta + c_4u_{t1} + c_6u_{11} + c_7u_{12} = 0
 \label{c4c6c7}
\end{equation}
where $\Delta$ is defined in \eqref{Del}.

\begin{Theorem}
  Equation \eqref{c4c6c7} is integrable with the Lax pair \eqref{Lax} and  recursion relations for symmetries \eqref{recurs}
where operators $A_i$ and $B_i$ are defined in \eqref{c4AiBi}.
\end{Theorem}

{\bf Proof:}
The symmetry condition becomes
\begin{equation}
 M[\varphi] = \{M_{12}D_t + M_{2t}D_1 + M_{t1}D_2 + c_4D_tD_1 + c_6D_1^2 + c_7D_1D_2\}\varphi  = 0.
 \label{symc4}
\end{equation}
It is identically satisfied by $\varphi = u_1$. Combining equation $M[u_1] = 0$ with the second equation from \eqref{ident}
we obtain
\begin{equation}
  M_{12}u_{t1} + M_{2t}u_{11} + M_{t1}u_{12} + D_1(c_4u_{t1} + c_6u_{11} + c_7u_{12}) = 0.
 \label{Mut1}
\end{equation}
We use \eqref{Mut1} in the identity transformation
\begin{eqnarray}
&& M_{2t}D_1 = M_{2t}u_{11}\frac{1}{u_{11}}D_1\nonumber\\
&& = - \{M_{12}u_{t1} + M_{t1}u_{12} + D_1(c_4u_{t1} + c_6u_{11} + c_7u_{12})\}\frac{1}{u_{11}}D_1.
\label{M2tD1}
\end{eqnarray}
Applying this to the symmetry condition \eqref{symc4} we transform it to the skew-factorized form \eqref{factor}
\begin{equation}
  \left\{(M_{12} + c_4D_1)\left(D_t - \frac{u_{t1}}{u_{11}}D_1\right) + (M_{t1} + c_7D_1)\left(D_2 - \frac{u_{12}}{u_{11}}D_1\right)\right\}\varphi = 0.
 \label{skewc4}
\end{equation}
We define
\begin{eqnarray}
& A_1 = \displaystyle\frac{1}{u_{11}}(M_{12} + c_4D_1),\quad A_2 = \frac{1}{u_{11}}(M_{t1} + c_7D_1)\nonumber\\
& B_1 = - D_2 \displaystyle + \frac{u_{12}}{u_{11}}D_1,\quad B_2 = D_t - \frac{u_{t1}}{u_{11}}D_1
 \label{c4AiBi}
\end{eqnarray}
so that \eqref{skewc4} takes the skew-factorized form \eqref{factor}.
A straightforward check shows that all the integrability conditions \eqref{commut} are identically satisfied on solutions of equation \eqref{c4c6c7}, together with their immediate consequences for the Lax pair \eqref{Lax} and  recursion relations for symmetries \eqref{recurs}.

The Lax pair and recursion relations for symmetries are obtained by using \eqref{c4AiBi} in the formulas \eqref{Lax} and \eqref{recurs}, respectively.
\hfill $\blacksquare$

\subsection{$(c_5c_7c_8)$-parameter system}

Now we consider the three-parameter equation obtained by setting $a_5 = c_4 = c_6 = c_9 = 0$ in \eqref{eqI}
\begin{equation}
  \Delta + c_5u_{t2} + c_7u_{12} + c_8u_{22} = 0
 \label{c5c7c8}
\end{equation}
where $\Delta$ is defined in \eqref{Del}.

\begin{Theorem}
   Equation \eqref{c5c7c8} is integrable with the Lax pair \eqref{Lax} and  recursion relations for symmetries \eqref{recurs}
where operators $A_i$ and $B_i$ are defined in \eqref{c5AiBi}.
\end{Theorem}

{\bf Proof:}
The symmetry condition becomes
\begin{equation}
 M[\varphi] = \{M_{12}D_t + M_{2t}D_1 + M_{t1}D_2 + c_5D_tD_2 + c_7D_1D_2 + c_8D_2^2\}\varphi  = 0.
 \label{symc5}
\end{equation}
It is identically satisfied by $\varphi = u_2$. We combine the equation $M[u_2] = 0$ with the second equation from \eqref{ident}
to obtain
\begin{equation}
  M_{12}u_{t2} + M_{2t}u_{12} + M_{t1}u_{22} + D_2(c_5u_{t2} + c_7u_{12} + c_8u_{22}) = 0.
 \label{Mut2}
\end{equation}
We use \eqref{Mut2} in the identity transformation
\begin{eqnarray}
&& M_{t1}D_2 = M_{t1}u_{22}\frac{1}{u_{22}}D_2\nonumber\\
&& = - \{M_{12}u_{t2} + M_{2t}u_{12} + D_2(c_5u_{t2} + c_7u_{12} + c_8u_{22})\}\frac{1}{u_{22}}D_2.
\label{Mt1D2}
\end{eqnarray}
Applying this to the symmetry condition \eqref{symc5} we transform it to the skew-factorized form \eqref{factor}
\begin{equation}
  \left\{(M_{12} + c_5D_2)\left(D_t - \frac{u_{t2}}{u_{22}}D_2\right) + (M_{2t} + c_7D_2)\left(D_1 - \frac{u_{12}}{u_{22}}D_2\right)\right\}\varphi = 0.
 \label{skewc5}
\end{equation}
We define
\begin{eqnarray}
& A_1 = \displaystyle\frac{1}{u_{22}}(M_{12} + c_5D_2),\quad A_2 = \frac{1}{u_{22}}(M_{2t} + c_7D_2)\nonumber\\
& B_1 = - D_1 \displaystyle + \frac{u_{12}}{u_{22}}D_2,\quad B_2 = D_t - \frac{u_{t2}}{u_{22}}D_2
 \label{c5AiBi}
\end{eqnarray}
so that \eqref{skewc5} takes the skew-factorized form \eqref{factor}.
A straightforward check shows that all the integrability conditions \eqref{commut} are identically satisfied on solutions of equation \eqref{c5c7c8}, together with their immediate consequences for the Lax pair \eqref{Lax} and  recursion relations for symmetries \eqref{recurs}.

The Lax pair and recursion relations for symmetries are obtained by using \eqref{c5AiBi} in the formulas \eqref{Lax} and \eqref{recurs}, respectively.
\hfill $\blacksquare$

\section{Notation for bi-Hamiltonian systems}
\setcounter{equation}{0}
\label{sec-biHam}

Here we present the notation convenient for bi-Hamiltonian systems and some remarks concerning second Hamiltonian operators.

\begin{equation}
\Box = u_{11}u_{22} - u_{12}^2 + a_5, \quad \Phi = v_1u_{22} - v_2u_{12},\quad \chi = v_1u_{12} - v_2u_{11}.
 \label{fixi}
\end{equation}
The determinant $\Delta$ defined in \eqref{Del} becomes
\begin{equation}
\Delta = v_t(\Box - a_5) - v_1\Phi + v_2\chi.
 \label{delt}
\end{equation}
We introduce the operators
\begin{equation}
\Psi = \Phi D_1 - \chi D_2, \Longrightarrow D_1\Phi - D_2\chi = - \Psi^T,\quad \hat{\Psi} = \Psi - c_4D_1 - c_5D_2
 \label{Psi}
\end{equation}
where $T$ denotes transposed operator. We will use the result
\begin{equation}
D_1[\Phi] - D_2[\chi] = v_{11}u_{22} + v_{22}u_{11} - 2 v_{12}u_{12}
 \label{D12}
\end{equation}
where square brackets denote values of operators.
We define the following operators
\begin{eqnarray}
&& \Gamma = v_2D_1 - v_1D_2,\quad \Upsilon = u_{12}D_1 - u_{11}D_2, \quad \Theta = u_{22}D_1 - u_{12}D_2\nonumber\\
&& \tilde{\Gamma} = a_5\Gamma + c_4\Upsilon + c_5\Theta,\quad \tilde{\Upsilon} = -(c_4\Gamma + c_6\Upsilon + c_7\Theta)\nonumber\\
&& \tilde{\Theta} = c_5\Gamma + c_7\Upsilon + c_8\Theta.
 \label{GaUpTh}
\end{eqnarray}
First-order differential operators \eqref{Mij} entering the symmetry condition operator \eqref{M} take the form
\begin{equation}
 M_{12} = (\Box - a_5) D_t - \Psi,\quad M_{2t} = - \Phi D_t + v_t \Theta - v_2\Gamma,\quad M_{t1} = \chi D_t - v_t \Upsilon + v_1 \Gamma.
 \label{M_ij}
\end{equation}

In the following sections, we only show that the second Hamiltonian operators $J_1=RJ_0$ are skew symmetric, $J_1^T = - J_1$.
A check of the Jacobi identities and compatibility of the two Hamiltonian operators $J_0$ and $J_1$ is straightforward but too much lengthy to be presented here.
These calculations are somewhat facilitated by P. Olver's method of functional multivectors \cite{olv}, chapter 7. Examples of such calculations can be found in our papers \cite{sym,sy}.

The resulting bi-Hamiltonian system has the form
\begin{equation}
 \left(
\begin{array}{c}
 u_t\\
 v_t
\end{array}
\right) = J_0
\left(
\begin{array}{c}
 \delta_u H_1\\
 \delta_v H_1
\end{array}
\right)
  = J_1
\left(
\begin{array}{c}
 \delta_u H_0\\
 \delta_v H_0
\end{array}
\right)
 \label{bi-Ham}
\end{equation}
where $J_0$ and $J_1$ are the first and second Hamiltonian operators, respectively, while $H_1$ and $H_0$ are the corresponding Hamiltonian densities.

\section{Summary of new bi-Hamiltonian systems}
\setcounter{equation}{0}
\label{RJ}

In this section we present new recursion and Hamiltonian operators and new bi-Hamiltonian (2+1)-dimensional systems in the form \eqref{bi-Ham}.
These results are given in the form of theorems while the appropriate proofs are transferred to the subsequent sections.

The first Hamiltonian operator \eqref{hamilton1} has the same general form for all new bi-Hamiltonian systems
\begin{equation}
    J_0 =
    \left(
    \begin{array}{rc}
   0, & \frac{1}{\Box}\\
  -\frac{1}{\Box}, & \frac{1}{\Box}(\hat{\Psi} - \Psi^T)\frac{1}{\Box}
    \end{array}
    \right)
\label{J0a5}
\end{equation}

\subsection{Bi-Hamiltonian form of $(a_5c_4c_5)$-parameter system}
\label{sec_a5c4c5}

According to \eqref{delt} the equation \eqref{a5c4c5} in the two-component form becomes
\begin{equation}
  u_t=v,\quad  v_t = \frac{v_1(\Phi - c_4) - v_2(\chi + c_5)}{\Box}\equiv \frac{(\Psi[v] - c_4v_1 - c_5v_2)}{\Box}\equiv \frac{\hat{\Psi}[v]}{\Box}.
 \label{vt a5c4c5}
\end{equation}

\begin{Theorem}
 Recursion operator for the system \eqref{vt a5c4c5} has the form
\begin{equation}
R = \left(\begin{array}{cc}
R_{11} & R_{12} \\
R_{21} & R_{22} \end{array}\right)
\label{R}
\end{equation}
where
\begin{eqnarray}
&& R_{11} = \tilde{\Gamma}^{-1}\hat{\Psi},\quad R_{12} = - \tilde{\Gamma}^{-1}\Box,\quad R_{22} = - \frac{1}{\Box}\Psi\tilde{\Gamma}^{-1}\Box \nonumber\\
&& R_{21} = \frac{1}{\Box(\Phi-c_4)}\{(\Phi-c_4)\Psi\tilde{\Gamma}^{-1}\hat{\Psi} - v_2\hat{\Psi} + \hat{\Psi}[v]D_2\}.
\label{Ra5}
\end{eqnarray}
\end{Theorem}
\begin{Theorem}
The second Hamiltonian operator has the form
\begin{equation}
 J_1 = RJ_0 = \left(
 \begin{array}{cc}
 \tilde{\Gamma}^{-1}, & \tilde{\Gamma}^{-1}\Psi^T\frac{1}{\Box} \\
 \frac{1}{\Box}\Psi \tilde{\Gamma}^{-1},& \frac{1}{\Box}(\Psi \tilde{\Gamma}^{-1} \Psi^T - \Gamma)\frac{1}{\Box}
 \end{array}\right).
 \label{J1a5}
\end{equation}
\end{Theorem}
\begin{Theorem}
The system \eqref{vt a5c4c5} has the bi-Hamiltonian form \eqref{bi-Ham} with operators $J_0$ and $J_1$ defined in \eqref{J0a5} and \eqref{J1a5}, respectively,
and the Hamiltonian densities $H_1 = v^2\Box/2$ and
\begin{equation}
 H_0 = \{F(v) + (c_5z_1 - c_4z_2)v\}(u_{11}u_{22} - u_{12}^2 + a_5)
 \label{H_0a5}
\end{equation}
where $F(v)$ is an arbitrary smooth function.
\end{Theorem}

\subsection{Bi-Hamiltonian form of $(c_4c_6c_7)$-parameter system}
\label{sec_c4c6c7}

According to \eqref{delt} the equation \eqref{c4c6c7} in the two-component form becomes
\begin{equation}
  u_t=v,\quad  v_t = \frac{(\Psi[v] - c_4v_1 - c_6u_{11} - c_7u_{12})}{\Box}\equiv \frac{\hat{\Psi}[v] - c_6u_{11} - c_7u_{12}}{\Box}.
 \label{vt c4c6c7}
\end{equation}

\begin{Theorem}
Recursion operator for the system \eqref{vt c4c6c7} has the form
\begin{equation}
R =
\left(\begin{array}{cc}
\tilde{\Upsilon}^{-1}\Psi & - \tilde{\Upsilon}^{-1}\Box \\
\frac{1}{\Box}(\hat{\Psi}\tilde{\Upsilon}^{-1}\Psi + \Upsilon) & - \frac{1}{\Box}\hat{\Psi}\tilde{\Upsilon}^{-1}\Box \end{array}\right).
\label{Rc4}
\end{equation}
\end{Theorem}
\begin{Theorem}
 The second Hamiltonian operator has  the form
 \begin{equation}
 J_1 = RJ_0 = \left(
 \begin{array}{cc}
 \tilde{\Upsilon}^{-1}, & \tilde{\Upsilon}^{-1}\hat{\Psi}^T\frac{1}{\Box} \\
 \frac{1}{\Box}\hat{\Psi} \tilde{\Upsilon}^{-1},& \frac{1}{\Box}\hat{(\Psi}\tilde{\Upsilon}^{-1}\hat{\Psi}^T + \Upsilon)\frac{1}{\Box}
 \end{array}\right).
 \label{J1c4}
\end{equation}
\end{Theorem}
\begin{Theorem}
The system \eqref{vt c4c6c7} has the bi-Hamiltonian form \eqref{bi-Ham}with operators $J_0$ and $J_1$ defined in \eqref{J0a5} and \eqref{J1c4}, respectively,
and the Hamiltonian densities
\begin{equation}
H_1 = \frac{v^2\Box}{2} + \frac{u}{2}(c_6u_{11} + c_7u_{12}).
 \label{H1c4}
\end{equation}
and
\begin{equation}
 H_0 = v\Box (c_6z_2 - c_7z_1 + f(u_1)) + c_4(F(u_1) + c_7u).
 \label{H_0c4}
\end{equation}
\end{Theorem}
Here $f(u_1)$ is an arbitrary smooth function and $F(u_1)$ is the antiderivative for $f(u_1)$.

\subsection{Bi-Hamiltonian form of $(c_5c_7c_8)$-parameter system}
\label{sec_c5c7c8}

According to \eqref{delt} the equation \eqref{c5c7c8} in the two-component form becomes
\begin{equation}
  u_t=v,\quad  v_t = \frac{(\Psi[v] - c_5v_2 - c_7u_{12} - c_8u_{22})}{\Box}\equiv \frac{\hat{\Psi}[v]  - c_7u_{12} - c_8u_{22} }{\Box}.
 \label{vt c5c7c8}
\end{equation}

\begin{Theorem}
 Recursion operator for the system \eqref{vt c5c7c8} has the form
\begin{equation}
R = \left(\begin{array}{cc}
\tilde{\Theta}^{-1}\Psi & - \tilde{\Theta}^{-1}\Box \\
\frac{1}{\Box}(\hat{\Psi}\tilde{\Theta}^{-1}\Psi - \Theta) & - \frac{1}{\Box}\hat{\Psi}\tilde{\Theta}^{-1}\Box \end{array}\right).
\label{Rc5}
\end{equation}
\end{Theorem}
\begin{Theorem}
 The second Hamiltonian operator has the form
\begin{equation}
 J_1 = RJ_0 = \left(
 \begin{array}{cc}
 \tilde{\Theta}^{-1}, & \tilde{\Theta}^{-1}\hat{\Psi}^T\frac{1}{\Box} \\
 \frac{1}{\Box}\hat{\Psi} \tilde{\Theta}^{-1},& \frac{1}{\Box}\hat{(\Psi}\tilde{\Theta}^{-1}\hat{\Psi}^T - \Theta)\frac{1}{\Box}
 \end{array}\right).
 \label{J1c5}
\end{equation}
\end{Theorem}
\begin{Theorem}
The system \eqref{vt c5c7c8} has the bi-Hamiltonian form \eqref{bi-Ham}with operators $J_0$ and $J_1$ defined in \eqref{J0a5} and \eqref{J1c5}, respectively,
and the Hamiltonian densities
\begin{equation}
H_1 = \frac{v^2\Box}{2} + \frac{u}{2}(c_7u_{12} + c_8u_{22}).
 \label{H1c5}
\end{equation}
and
\begin{equation}
 H_0 = v\Box (c_8z_1 - c_7z_2 + f(u_2)) + c_5(F(u_2) + c_7u).
 \label{H_0c5}
\end{equation}
\end{Theorem}
Here $F(u_2)$ is the antiderivative of $f(u_2)$, the latter being an arbitrary smooth function.

\section{Bi-Hamiltonian form of $(a_5c_4c_5)$-parameter system: proofs}
\setcounter{equation}{0}
\label{sec-a5c4c5}

The equation \eqref{a5c4c5} in the two-component form is given in \eqref{vt a5c4c5}.

We will use the inverse operator $\tilde{\Gamma}^{-1}$ which can make sense merely as a \textit{formal} inverse. Thus, the relations involving $\tilde{\Gamma}^{-1}$ are also formal. The proper interpretation of the inverse operators and relations involving them requires the language of differential coverings (see the original papers \cite{Gu,Marvan} and the recent survey \cite{Kras}).

We specify the inverse of $\tilde{\Gamma}$ by the property $\tilde{\Gamma}^{-1}\tilde{\Gamma} = I$ where $I$ is the unit (identity) operator. A detailed example of constructing such an inverse operator was given in our paper \cite{sym}.

According to the definitions \eqref{a5AiBi} and \eqref{M_ij}, we have
\begin{eqnarray}
&& A_1 = \frac{1}{v_t}\{- (\Phi - c_4)D_t + v_t\Theta - v_2\Gamma\},\quad B_1 =  \frac{1}{v_t}(v_2D_t - v_tD_2)\nonumber\\
&& A_2 = \frac{1}{v_t}\{(\chi + c_5)D_t - v_t\Upsilon + v_1\Gamma\},\quad B_2 =  \frac{1}{v_t}(v_tD_1 - v_1D_t).\nonumber\\
 \label{a5AB}
\end{eqnarray}
Recursion relations \eqref{recurs} become
\begin{eqnarray}
&& - (\Phi - c_4)\tilde{\psi} + (v_t\Theta - v_2\Gamma)\tilde{\varphi} = v_2\psi - v_tD_2\varphi\nonumber\\
&& (\chi + c_5) \tilde{\psi} + (- v_t\Upsilon + v_1\Gamma)\tilde{\varphi} = v_tD_1\varphi - v_1\psi
 \label{recurs_a5}
\end{eqnarray}
where $\varphi$ and $\tilde{\varphi}$ are symmetry characteristics for the original and transformed symmetry, respectively, and $\psi=\varphi_t$,
$\tilde{\psi} = \tilde{\varphi}_t$. The subscripts denote partial derivatives. Combining the two equations in \eqref{recurs_a5} we eliminate $\tilde{\psi}$ with the result
\begin{equation}
\tilde{\Gamma}\tilde{\varphi} = \hat{\Psi}\varphi - \Box\psi \iff \tilde{\varphi} = \tilde{\Gamma}^{-1}(\hat{\Psi}\varphi - \Box\psi).
 \label{fi_recurs_a5}
\end{equation}
Utilization of \eqref{fi_recurs_a5} in \eqref{recurs_a5} yields only one independent equation
\begin{eqnarray}
&& \tilde{\psi} = \frac{1}{\Box(\Phi-c_4)}\{(\Phi-c_4)\Psi\tilde{\Gamma}^{-1}\hat{\Psi} - v_2\hat{\Psi} + \hat{\Psi}[v]D_2\}\varphi \nonumber\\
&& - \frac{1}{\Box}\Psi\tilde{\Gamma}^{-1}\Box\psi
\label{ps_recurs_a5}
\end{eqnarray}
where we have used the relations
\[v_1\Theta - v_2\Upsilon = \Psi,\quad \hat{\Psi}[v]\Theta - \Box v_2\Gamma = (\Phi - c_4)\Psi - v_2\tilde{\Gamma}.\]
Recursion relations \eqref{fi_recurs_a5} and \eqref{ps_recurs_a5} can be written in the form of a matrix recursion operator $R$
\begin{equation*}
\left(\begin{array}{c}
\tilde{\varphi}\\ \tilde{\psi}\end{array}\right) = R\left(\begin{array}{c} \varphi\\ \psi \end{array}\right) =
\left(\begin{array}{cc}
R_{11} & R_{12} \\
R_{21} & R_{22} \end{array}\right) \left(\begin{array}{c} \varphi\\ \psi \end{array}\right)
% \label{R}
\end{equation*}
with the matrix elements \eqref{Ra5}.

The first Hamiltonian operator \eqref{hamilton1} for equation \eqref{a5c4c5} due to \eqref{K11} is determined by \eqref{J0a5}
and the corresponding Hamiltonian density \eqref{H1} becomes
\begin{equation}
H_1 = v^2\Box/2.
 \label{H1a5}
\end{equation}

The second Hamiltonian operator obtained by the formula $J_1=RJ_0$ has the form \eqref{J1a5}
The operator $J_1$ in \eqref{J1a5} is manifestly skew symmetric, same as $J_0$ in \eqref{J0a5}.

The remaining task is to find the Hamiltonian density $H_0$ corresponding to the new Hamiltonian operator \eqref{J1a5} according to the formula
\begin{equation}
J_1
\left(
\begin{array}{c}
 \delta_u H_0\\
 \delta_v H_0
\end{array}
\right) =
\left(
\begin{array}{c}
 v\\[2pt]
\displaystyle \frac{\hat{\Psi}[v]}{\Box}
\end{array}
\right)
  \label{Ham_a5}
  \end{equation}
where \eqref{vt a5c4c5} has been used. We assume that $H_0$ does not depend on partial derivatives of $v$, so that $\delta_vH_0 = H_{0,v}$.

The first line of equation \eqref{Ham_a5} with $J_1$ defined in \eqref{J1a5}
\begin{equation}
\tilde{\Gamma}^{-1}\left(\delta_uH_0 + \Psi^T\frac{H_{0,v}}{\Box}\right) = v \iff \delta_uH_0 = - \Psi^T\frac{H_{0,v}}{\Box} + c_4\chi + c_5\Phi
\label{1stline}
\end{equation}
being used in the second line of \eqref{Ham_a5}
\[\frac{1}{\Box}\Psi\tilde{\Gamma}^{-1}\left(\delta_uH_0 + \Psi^T\frac{H_{0,v}}{\Box}\right) - \frac{1}{\Box}\Gamma\frac{H_{0,v}}{\Box}
= \frac{\hat{\Psi}[v]}{\Box}\]
implies $\Gamma[H_{0,v}/\Box] = c_4v_1 + c_5v_2 \iff v_2D_1[H_{0,v}/\Box] - v_1D_2[H_{0,v}/\Box] = c_4v_1 + c_5v_2$. This equation implies
\begin{equation}
 \frac{H_{0,v}}{\Box} = c_5z_1 - c_4z_2 + f(v) \iff H_0 = \Box \{(c_5z_1 - c_4z_2)v + F(v)\} + h[u]
 \label{H0}
\end{equation}
where $f(v)$ is an arbitrary smooth function belonging to the kernel of $\Gamma$, $F$ is the antiderivative for $f$,
$\Box = u_{11}u_{22} - u_{12}^2 + a_5$ and $h[u]$ is a function only of $u$ and its partial derivatives in $z_1,z_2$.

$H_0$ in \eqref{H0} should satisfy the second equation in \eqref{1stline}
which yields
\begin{eqnarray}
 \delta_uH_0 &=& (c_5z_1 - c_4z_2 + f(v))(v_{11}u_{22} + v_{22}u_{11} - 2v_{12}u_{12})\nonumber\\
&+& f'(v)(v_1^2u_{22} + v_2^2u_{11} - 2v_1v_2u_{12}) + 2(c_4\chi + c_5\Phi).
 \label{deluH0a5}
\end{eqnarray}
Calculating directly the variational derivative $\delta_uH_0$ from $H_0$ in \eqref{H0} and comparing it with \eqref{deluH0a5} we obtain $\delta_uh[u]=0$.
In the final result we skip the "null Hamiltonian" $h[u]$  and obtain the result \eqref{H_0a5} for $H_0$.

Thus, bi-Hamiltonian representation of the $(a_5c_4c_5)$-parameter system \eqref{vt a5c4c5} has the form \eqref{bi-Ham} with $J_0$ defined in \eqref{J0a5}, $H_1$ in \eqref{H1a5}, $J_1$ in \eqref{J1a5}, $H_0$ in \eqref{H_0a5} and the recursion operator $R$ determined by \eqref{Ra5}.

\section{Bi-Hamiltonian form of $(c_4c_6c_7)$-parameter system: proofs}
\setcounter{equation}{0}
\label{sec-c4c6c7}

The equation \eqref{c4c6c7} in the two-component form is given in \eqref{vt c4c6c7}.

We will use the inverse operator $\tilde{\Upsilon}^{-1}$ which we specify by the property $\tilde{\Upsilon}^{-1}\tilde{\Upsilon} = I$.
According to the definitions \eqref{c4AiBi} and \eqref{M_ij}, we have
\begin{eqnarray}
&& A_1 = \frac{1}{u_{11}}(\Box D_t - \Psi +c_4D_1),\quad B_1 =  \frac{1}{u_{11}}\Upsilon\equiv \frac{1}{u_{11}}(u_{12}D_1 - u_{11}D_2)\nonumber\\
&& A_2 = \frac{1}{u_{11}}\{\chi D_t - v_t\Upsilon + v_1\Gamma + c_7D_1\},\quad B_2 =  \frac{1}{u_{11}}(u_{11}D_t - v_1D_1).\nonumber\\
 \label{c4AB}
\end{eqnarray}
Recursion relations \eqref{recurs} become
\begin{eqnarray}
&& \Box\tilde{\psi} - \hat{\Psi}\tilde{\varphi} = \Upsilon\varphi\nonumber\\
&& \chi\tilde{\psi} + (- v_t\Upsilon + v_1\Gamma + c_7D_1)\tilde{\varphi} = - v_1D_1\varphi + u_{11}\psi
 \label{recurs_c4}
\end{eqnarray}
where $\psi=\varphi_t$ and $\tilde{\psi} = \tilde{\varphi}_t$. Combining the two equations in \eqref{recurs_c4} we eliminate $\tilde{\psi}$ with the result
\begin{equation}
\tilde{\Upsilon}\tilde{\varphi} = \Psi\varphi - \Box\psi \iff \tilde{\varphi} = \tilde{\Upsilon}^{-1}(\Psi\varphi - \Box\psi).
 \label{fi_recurs_c4}
\end{equation}
Utilization of \eqref{fi_recurs_c4} in \eqref{recurs_c4} yields only one independent equation
\begin{equation}
 \tilde{\psi} = \frac{1}{\Box}(\hat{\Psi}\tilde{\Upsilon}^{-1}\Psi + \Upsilon)\varphi - \frac{1}{\Box}\hat{\Psi}\tilde{\Upsilon}^{-1}\Box\psi .
\label{ps_recurs_c4}
\end{equation}
Recursion relations \eqref{fi_recurs_c4} and \eqref{ps_recurs_c4} can be written in the form of a matrix recursion operator $R$
\begin{equation*}
\left(\begin{array}{c}
\tilde{\varphi}\\ \tilde{\psi}\end{array}\right) = R\left(\begin{array}{c} \varphi\\ \psi \end{array}\right) =
\left(\begin{array}{cc}
\tilde{\Upsilon}^{-1}\Psi & - \tilde{\Upsilon}^{-1}\Box \\
\frac{1}{\Box}(\hat{\Psi}\tilde{\Upsilon}^{-1}\Psi + \Upsilon) & - \frac{1}{\Box}\hat{\Psi}\tilde{\Upsilon}^{-1}\Box \end{array}\right)
\left(\begin{array}{c} \varphi\\ \psi \end{array}\right).
% \label{Rc4}
\end{equation*}

The first Hamiltonian operator \eqref{hamilton1} for equation \eqref{c4c6c7} due to \eqref{K11} takes the form \eqref{J0a5}
and the corresponding Hamiltonian density \eqref{H1} becomes
\begin{equation*}
H_1 = \frac{v^2\Box}{2} + \frac{u}{2}(c_6u_{11} + c_7u_{12}).
% \label{H1c4}
\end{equation*}

The second Hamiltonian operator obtained by the formula $J_1=RJ_0$ has the form \eqref{J1c4}.
The operator $J_1$ in \eqref{J1c4} is manifestly skew symmetric, same as $J_0$ in \eqref{J0a5}.

The remaining task is to find the Hamiltonian density $H_0$ corresponding to the new Hamiltonian operator \eqref{J1c4} according to the formula
\begin{equation}
J_1
\left(
\begin{array}{c}
 \delta_u H_0\\
 \delta_v H_0
\end{array}
\right) =
\left(
\begin{array}{c}
 v\\[2pt]
\displaystyle \frac{1}{\Box}\left(\hat{\Psi}[v] - c_6u_{11} - c_7u_{12}\right)
\end{array}
\right)
 \label{Ham_c4}
  \end{equation}
where \eqref{vt c4c6c7} has been used. We assume that $H_0$ does not depend on partial derivatives of $v$, so that $\delta_vH_0 = H_{0,v}$.
The first line of equation \eqref{Ham_c4} with $J_1$ defined in \eqref{J1c4}
\begin{equation}
\tilde{\Upsilon}^{-1}\left(\delta_uH_0 + \hat{\Psi}^T\frac{H_{0,v}}{\Box}\right) = v \iff \delta_uH_0 = - \hat{\Psi}^T\frac{H_{0,v}}{\Box}
- (c_6\Upsilon[v] + c_7\Theta[v])
\label{1stc4}
\end{equation}
being used in the second line of \eqref{Ham_c4}
\[\frac{1}{\Box}\hat{\Psi}\tilde{\Upsilon}^{-1}\left(\delta_uH_0 + \hat{\Psi}^T\frac{H_{0,v}}{\Box}\right) + \frac{1}{\Box}\Upsilon\frac{H_{0,v}}{\Box}
= \frac{1}{\Box}(\hat{\Psi}[v] - c_6u_{11} - c_7u_{12})\]
implies $\Upsilon[H_{0,v}/\Box] = - (c_6u_{11} + c_7u_{12}) \iff H_{0,v}/\Box = c_6z_2 - c_7z_1 + f(u_1)$ where $f$ is an arbitrary smooth function belonging to the kernel of $\Upsilon$. This equation implies
\begin{equation}
 H_0 = v\Box (c_6z_2 - c_7z_1 + f(u_1)) + h[u]
 \label{H0c4}
\end{equation}
where $h[u]$ is a function only of $u$ and its partial derivatives in $z_1,z_2$.
$H_0$ in \eqref{H0c4} should satisfy the second equation in \eqref{1stc4}
which yields
\begin{eqnarray}
 \delta_uH_0 &=& (c_6z_2 - c_7z_1 + f(u_1))(v_{11}u_{22} + v_{22}u_{11} - 2v_{12}u_{12})\nonumber\\
&+& v_1(f'(u_1)\Box - 2c_6u_{12} - 2c_7u_{22}) + 2v_2(c_6u_{11} +c_7u_{12})\nonumber\\
&-& c_4f'(u_1)u_{11} + c_4c_7.
 \label{deluH0c4}
\end{eqnarray}
Calculating directly the variational derivative $\delta_uH_0$ from $H_0$ in \eqref{H0c4} and comparing it with \eqref{deluH0c4} we obtain
$\delta_uh[u]=c_4(-f'(u_1)u_{11} + c_7)$. Introducing $F(u_1)$ to be the antiderivative of $f(u_1)$ we finally obtain the result \eqref{H_0c4} for $H_0$.

Thus, bi-Hamiltonian representation of the $(c_4c_6c_7)$-parameter system \eqref{vt c4c6c7} has the form \eqref{bi-Ham} with $J_0$ defined in \eqref{J0a5}, $H_1$ in \eqref{H1c4}, $J_1$ in \eqref{J1c4}, $H_0$ in \eqref{H_0c4} and the recursion operator $R$ determined in \eqref{Rc4}.

 \section{Bi-Hamiltonian form of $(c_5c_7c_8)$-parameter system: proofs}
\setcounter{equation}{0}
\label{sec-c5c7c8}

We will use the inverse operator $\tilde{\Theta}^{-1}$ which we specify by the property $\tilde{\Theta}^{-1}\tilde{\Theta} = I$.

According to the definitions \eqref{c5AiBi} and \eqref{M_ij}, we have
\begin{eqnarray}
&& A_1 = \frac{1}{u_{22}}(\Box D_t - \Psi + c_5D_2),\quad B_1 = - \frac{1}{u_{22}}\Theta\equiv \frac{1}{u_{22}}(u_{12}D_2 - u_{22}D_1)\nonumber\\
&& A_2 = \frac{1}{u_{22}}\{- \Phi D_t + v_t\Theta - v_2\Gamma + c_7D_2\},\quad B_2 =  \frac{1}{u_{22}}(u_{22}D_t - v_2D_2).\nonumber\\
 \label{c5 AB}
\end{eqnarray}
Recursion relations \eqref{recurs} become
\begin{eqnarray}
&& \Box\tilde{\psi} - \hat{\Psi}\tilde{\varphi} = - \Theta\varphi\nonumber\\
&& - \Phi\tilde{\psi} + (v_t\Theta - v_2\Gamma + c_7D_2)\tilde{\varphi} = - v_2D_2\varphi + u_{22}\psi
 \label{recurs_c5}
\end{eqnarray}
where $\psi=\varphi_t$ and $\tilde{\psi} = \tilde{\varphi}_t$. Combining the two equations in \eqref{recurs_c5} we eliminate $\tilde{\psi}$ with the result
\begin{equation}
\tilde{\Theta}\tilde{\varphi} = \Psi\varphi - \Box\psi \iff \tilde{\varphi} = \tilde{\Theta}^{-1}(\Psi\varphi - \Box\psi).
 \label{fi_recurs_c5}
\end{equation}
Utilization of \eqref{fi_recurs_c5} in \eqref{recurs_c5} yields only one independent equation
\begin{equation}
 \tilde{\psi} = \frac{1}{\Box}(\hat{\Psi}\tilde{\Theta}^{-1}\Psi - \Theta)\varphi - \frac{1}{\Box}\hat{\Psi}\tilde{\Theta}^{-1}\Box\psi .
\label{ps_recurs_c5}
\end{equation}
Recursion relations \eqref{fi_recurs_c5} and \eqref{ps_recurs_c5} can be written in the form of a matrix recursion operator $R$
\begin{equation*}
\left(\begin{array}{c}
\tilde{\varphi}\\ \tilde{\psi}\end{array}\right) = R\left(\begin{array}{c} \varphi\\ \psi \end{array}\right) =
\left(\begin{array}{cc}
\tilde{\Theta}^{-1}\Psi & - \tilde{\Theta}^{-1}\Box \\
\frac{1}{\Box}(\hat{\Psi}\tilde{\Theta}^{-1}\Psi - \Theta) & - \frac{1}{\Box}\hat{\Psi}\tilde{\Theta}^{-1}\Box \end{array}\right)
\left(\begin{array}{c} \varphi\\ \psi \end{array}\right).
% \label{Rc5}
\end{equation*}

The first Hamiltonian operator \eqref{hamilton1} for equation \eqref{c5c7c8} due to \eqref{K11} takes the form
\begin{equation}
    J_0 =
    \left(
    \begin{array}{rc}
   0, & \frac{1}{\Box}\\
  -\frac{1}{\Box}, & \frac{1}{\Box}(\hat{\Psi} - \Psi^T)\frac{1}{\Box}
    \end{array}
    \right)
\label{J0c5}
\end{equation}
and the corresponding Hamiltonian density \eqref{H1} becomes
\begin{equation*}
H_1 = \frac{v^2\Box}{2} + \frac{u}{2}(c_7u_{12} + c_8u_{22}).
% \label{H1c5}
\end{equation*}

The second Hamiltonian operator obtained by the formula $J_1=RJ_0$ has the form
\begin{equation*}
 J_1 = \left(
 \begin{array}{cc}
 \tilde{\Theta}^{-1}, & \tilde{\Theta}^{-1}\hat{\Psi}^T\frac{1}{\Box} \\
 \frac{1}{\Box}\hat{\Psi} \tilde{\Theta}^{-1},& \frac{1}{\Box}\hat{(\Psi}\tilde{\Theta}^{-1}\hat{\Psi}^T - \Theta)\frac{1}{\Box}
 \end{array}\right).
% \label{J1c5}
\end{equation*}
The operator $J_1$ in \eqref{J1c5} is manifestly skew symmetric, same as $J_0$ in \eqref{J0c5}.

The remaining task is to find the Hamiltonian density $H_0$ corresponding to the new Hamiltonian operator \eqref{J1c5} according to the formula
\begin{equation}
J_1
\left(
\begin{array}{c}
 \delta_u H_0\\
 \delta_v H_0
\end{array}
\right) =
\left(
\begin{array}{c}
 v\\[2pt]
\displaystyle \frac{1}{\Box}\left(\hat{\Psi}[v] - c_7u_{12} - c_8u_{22}\right)
\end{array}
\right)
  \label{Ham_c5}
  \end{equation}
where \eqref{vt c5c7c8} has been used. We assume that $H_0$ does not depend on partial derivatives of $v$, so that $\delta_vH_0 = H_{0,v}$.
The first line of equation \eqref{Ham_c5} with $J_1$ defined in \eqref{J1c5}
\begin{equation}
\tilde{\Theta}^{-1}\left(\delta_uH_0 + \hat{\Psi}^T\frac{H_{0,v}}{\Box}\right) = v \iff \delta_uH_0 = - \hat{\Psi}^T\frac{H_{0,v}}{\Box}
+ c_7\chi + c_8\Phi
\label{1stc5}
\end{equation}
being used in the second line of \eqref{Ham_c5}
\[\frac{1}{\Box}\hat{\Psi}\tilde{\Theta}^{-1}\left(\delta_uH_0 + \hat{\Psi}^T\frac{H_{0,v}}{\Box}\right) - \frac{1}{\Box}\Theta\frac{H_{0,v}}{\Box}
= \frac{1}{\Box}(\hat{\Psi}[v] - c_7u_{12} - c_8u_{22})\]
implies $\Theta[H_{0,v}/\Box] = c_7u_{12} + c_8u_{22} \iff H_{0,v}/\Box = c_8z_1 - c_7z_2 + f(u_2)$ where $f$ is an arbitrary smooth function belonging to the kernel of $\Theta$. This equation implies
\begin{equation}
 H_0 = v\Box (c_8z_1 - c_7z_2 + f(u_2)) + h[u]
 \label{H0c5}
\end{equation}
where $h[u]$ is a function only of $u$ and its partial derivatives in $z_1,z_2$.
$H_0$ in \eqref{H0c5} should satisfy the second equation in \eqref{1stc5} %\eqref{1stline}
which yields
\begin{eqnarray}
 \delta_uH_0 &=& (c_8z_1 - c_7z_2 + f(u_2))(v_{11}u_{22} + v_{22}u_{11} - 2v_{12}u_{12})\nonumber\\
&+& 2c_7\chi + 2c_8\Phi + v_2f'(u_2)\Box + c_5(c_7 - f'(u_2)u_{22}).
 \label{deluH0c5}
\end{eqnarray}
Calculating directly the variational derivative $\delta_uH_0$ from $H_0$ in \eqref{H0c5} and comparing it with \eqref{deluH0c5} we obtain
$\delta_uh[u]=c_5(-f'(u_2)u_{22} + c_7)$. Introducing $F(u_2)$ to be the antiderivative of $f(u_2)$ we finally obtain
\begin{equation*}
 H_0 = v\Box (c_8z_1 - c_7z_2 + f(u_2)) + c_5(F(u_2) + c_7u).
% \label{H_0c5}
\end{equation*}

Thus, bi-Hamiltonian representation of the $(c_5c_7c_8)$-parameter system \eqref{vt c5c7c8} has the form \eqref{bi-Ham} with $J_0$ defined in \eqref{J0c5}, $H_1$ in \eqref{H1c5}, $J_1$ in \eqref{J1c5}, $H_0$ in \eqref{H_0c5} and the recursion operator $R$ determined in \eqref{Rc5}.

\section*{Conclusion}

We have obtained the general form of Euler-Lagrange evolutionary equations in $(2+1)$ dimensions containing only second order partial derivatives of the unknown.
Their Lagrangians have also been constructed. We have converted these equations into two-component evolutionary form and obtained Lagrangians for the two-component systems.
The Lagrangians are degenerate because the momenta cannot be inverted for the velocities. Applying to these degenerate Lagrangians the Dirac's theory of constraints, we have obtained a symplectic operator and its inverse, the Hamiltonian operator $J_0$ for each such system together with the Hamiltonian density $H_1$. Thus, all these systems have been presented in a Hamiltonian form.

We have explicitly demonstrated how the presentation of a symmetry condition in the skew-factorized form supply Lax pairs and recursion relations without the previous knowledge of Lax pairs.
In particular, we have shown how the symmetry condition for three-parameter cubic equations can be converted to a skew-factorized form and obtained Lax pair and recursion relations for such an equation. This procedure may serve as a hint for a future general method for skew-factorization of the symmetry condition.

We have derived recursion operators in a $2\times 2$ matrix form for the three-parameter two-component cubic systems. Composing the recursion operators $R$ with the Hamiltonian operator $J_0$ we have obtained second Hamiltonian operators $J_1=RJ_0$ for all such systems. We have found the Hamiltonian density $H_0$ corresponding to $J_1$, thus ending up with three new bi-Hamiltonian three-parameter cubic systems in $(2+1)$ dimensions.

\section*{Acknowledgments}

The authors are grateful to an anonymous reviewer for his important comments and suggestions which hopefully improved our paper.

The research of D. Yaz{\i}c{\i} is supported by the Research Fund of Y{\i}ld{\i}z Technical University, Turkey. Project Number: 4464.

\end{document}